\documentclass[authoryear,12pt,3p]{jowarticle}
\usepackage{graphicx}
\usepackage{amsmath}
\usepackage{amssymb}
\usepackage{natbib}
\usepackage{setspace}
\usepackage[all]{xy}
\usepackage{enumitem}
\usepackage{titlesec}
\makeatletter
\renewcommand\section{\@startsection{section}{1}{\z@}{-3.25ex plus -1ex minus -.2ex}{1.5ex plus .2ex}{\normalsize\bf}}
\renewcommand\subsection{\@startsection{subsection}{2}{\z@}{-3.25ex plus -1ex minus -.2ex}{1.5ex plus .2ex}{\normalsize\bf}}
\renewcommand\subsubsection{\@startsection{subsubsection}{3}{\z@}{-3.25ex plus -1ex minus -.2ex}{1.5ex plus .2ex}{\normalsize\bf}}
\makeatother

\begin{document}
\begin{frontmatter}
\title{On Gravitational Energy in Newtonian Theories}
\author{Neil Dewar}
\address{Munich Center for Mathematical Philosophy\\ LMU Munich, Germany}
\author{James Owen Weatherall}\ead{weatherj@uci.edu}
\address{Department of Logic and Philosophy of Science\\ University of California, Irvine, USA}
\begin{abstract}There are well-known problems associated with the idea of (local) gravitational energy in general relativity.  We offer a new perspective on those problems by comparison with Newtonian gravitation, and particularly geometrized Newtonian gravitation (i.e., Newton-Cartan theory).  We show that there is a natural candidate for the energy density of a Newtonian gravitational field.  But we observe that this quantity is gauge dependent, and that it cannot be defined in the geometrized (gauge-free) theory without introducing further structure.  We then address a potential response by showing that there is an analogue to the Weyl tensor in geometrized Newtonian gravitation.\end{abstract}
\end{frontmatter}
\doublespacing
\section{Introduction}\label{introduction}

There is a strong and natural sense in which, in general relativity, there are metrical degrees of freedom---including purely gravitational, i.e., source-free degrees of freedom, such as gravitational waves---that can influence the behavior of other physical systems.  And yet, if one attempts to associate a notion of local energy density with these metrical degrees of freedom, one quickly encounters (well-known) problems.\footnote{See, for instance, \citet[pp. 466-468]{MTW}, for a classic argument that there cannot be a local notion of energy associated with gravitation in general relativity; see also \citet{Choquet-Brouhat}, \citet{Goldberg}, and \citet{Curiel} for other arguments and discussion.}

For instance, consider the following simple argument that there cannot be a smooth tensor field representing gravitational energy-momentum in general relativity.\footnote{One of us (Weatherall) learned of this argument from David Malament, who attributes it to Bob Geroch; it is also described by Erik \citet{Curiel}, who likewise heard of it from Malament.}  Fix a relativistic spacetime $(M,g_{ab})$ and suppose there were a gravitational energy-momentum tensor, $T^G_{ab}$, associated with it.  Then this energy-momentum tensor either appears along with the rest of the energy-momentum content of the universe on the right-hand side of Einstein's equation,
\[
R_{ab} - \frac{1}{2}g_{ab} R = 8\pi T^{\text{Total}}_{ab};
\]
or else it does not appear.\footnote{Here $R^a{}_{bcd}$ is the Riemann tensor associated with $g_{ab}$, $R_{ab}=R^n{}_{abn}$ is the Ricci tensor, and $R=R^a{}_a$ is the curvature scalar.  For details and background, consult \citet{MTW}, \citet{Wald}, or \citet{MalamentGR}.  (Our conventions follow \citet{MalamentGR}, and in particular, we work in geometric units where $c=G=1$.)}

Suppose it does appear.  In that case, it would have to vanish whenever the spacetime is Ricci-flat ($R_{ab}=\mathbf{0}$), since in that case, the left-hand side of Einstein's equation would vanish.\footnote{Of course, one could consider situations in which there is matter whose energy-momentum precisely cancel the gravitational energy-momentum.  But one can likewise consider situations in which there is no such matter present, and then the argument follows.}  But there are Ricci-flat spacetimes that have non-vanishing Riemann curvature, including spacetimes describing black holes and gravitational waves.  It would follow that, in general, there would be no gravitational energy-momentum associated with precisely those ``purely gravitational'' phenomena whose energy content we aimed to describe.\footnote{By ``purely gravitational'' phenomena, we mean only that these are phenomena that may exist in the absence of matter, or which may vary for fixed matter distribution; we do not mean to identify these phenomena with a ``gravitational field'' \citep{Lehmkuhl2008}.  We are grateful to an anonymous referee for urging us to clarify this point. \label{gravField}}  And so, whatever else is the case, $T^G_{ab}$ does not capture the energy-momentum content of gravitational degrees of freedom that led us to consider the question of gravitational energy in the first place.

Conversely, suppose $T^G_{ab}$ does not appear on the right-hand side of Einstein's equation.  It would follow, from the contracted Bianchi identity, that the tensor representing the total non-gravitational energy-momentum  content of the universe is divergence-free.  This result is naturally interpreted as capturing a notion of ``local conservation'', i.e., it implies that there is no local exchange of energy-momentum between the sum total of the sources appearing on the right-hand side of Einstein's equation and the gravitational degrees of freedom associated with $T^G_{ab}$.  In other words, gravitational energy-momentum would not be fungible with the energy-momentum of matter, and so the influence of gravitation on other degrees of freedom could not be explained via the (local) exchange of energy-momentum.  Once again, it would seem that $T^G_{ab}$ is unable to capture the phenomena we were interested in in the first place.

There are several ways to respond to this situation.  One response is to give up on the idea of gravitational energy altogether---or even to argue that our classical expectations concerning energy and energy conservation are simply not met in general relativity.  There is much to be said for this attitude, though it should be adopted with caution, since various notions of energy continue to be of significant pragmatic value, even in general relativity.  In this latter vein, it turns out that in some special cases, one \emph{can} introduce various notions of gravitational energy in general relativity.  For instance, in the presence of certain boundary conditions and global conditions, one can introduce \emph{global} notions of gravitational energy that associate a quantity of energy with entire slices of spacetime, such as the the ADM energy defined at spatial infinity \citep{ADM1, ADM2}, or the Bondi energy defined at null infinity \citep{BondiEnergy}.\footnote{One can also define various notions of \emph{quasi-local} energy-momentum, such as the Hawking energy \citep{HawkingEnergy} and the Geroch energy \citep{GerochEnergy}.  These generally make sense relative to some choice of 2-sphere embedded in a spacelike hypersurface.  See \citet{quasilocalReview} for a detailed discussion of the various options.}  Alternatively, one can introduce frame-dependent notions of gravitational energy, such as the so-called Einstein pseudotensor or the Landau-Lifshitz pseudotensor \citep[Ch. 11]{Landau+Lifshitz}.  But generally, expressions for gravitational energy in general relativity all suffer from a basic defect: they cannot be defined in general or without reference to some further, arbitrarily chosen and presumably unphysical, structure.\footnote{See \citet{lam2011gravitational} for a more detailed discussion of the role of background structures in defining gravitational energy in general relativity.  Note that in some cases, one can argue that the extra structure necessary to define gravitational energy has a physical interpretation, related to (for instance) some class of observers.  But it does not follow that there is some well-defined, observer-independent quantity that is being observed from various perspectives; to the contrary, what is captured by the available notions of gravitational energy is better understood as relative to the chosen structures.  Some authors embrace this: \citet{pitts2010gauge}, for instance, has argued that there are infinitely many gravitational energies in general relativity, and that one might as well take the collection of all of them to be the appropriate ``covariant'' characterization of gravitational energy in the theory.}

We will not discuss these proposals in more detail.  Instead, our purpose is to address a lingering feeling that there is something unsatisfactory about this situation.  This dissatisfaction stems, at least in part, from the following worries.  First, one might think that there \emph{is} a perfectly good notion of (local) energy density associated with a gravitational field in Newtonian gravitation.  Indeed, \citet{MaxwellEM} proposed such a quantity, in analogy with the electrostatic case; this quantity was later derived in detail and analyzed by \citet{Synge}, who also proposed an expression for the Newtonian gravitational stress and momentum.\footnote{See also \citet{quasilocalReview}.}  The existence of such quantities in Newtonian gravitation might well lead one to expect similar quantities to exist in general relativity---all the more so because, arguably, the gravitational degrees of freedom in general relativity are richer than in Newtonian gravitation.

But as we will argue in what follows, the situation concerning gravitational energy in Newtonian gravitation is not nearly so clear as this gloss suggests.  First, although one can introduce a notion of gravitational energy, and even a gravitational mass-momentum tensor, in Newtonian gravitation, there is a strong sense in which one can do so only relative to some further background structure.  There does not appear to be any suitably ``invariant'' notion of gravitational energy.  Moreover, once one moves to the geometrized reformulation of Newtonian gravitation, known as Newton-Cartan theory, the gravitational energy and mass-momentum tensors cannot be defined---even though, as we show below, this theory, too, has purely gravitational degrees of freedom.  All of this suggests that the non-existence of (local) gravitational energy is not special to general relativity, but is rather endemic to suitably ``geometrical'' theories---an observation that leads us to re-examine the conceptual underpinnings of the very notion of ``energy''.\footnote{Note the parallel here to previous arguments about the ways in which geometrized Newtonian gravitation can inform our understanding of the foundations of general relativity, concerning, for instance, the character of spacetime singularities \citep{WeatherallSingularity}; the status of inertial motion \citep{WeatherallSGP,WeatherallPuzzleball,WeatherallConservation}; the conventionality of geometry \citep{Manchak+Weatherall}; classical analogues of exact solutions of Einstein's equation \citep{EhlersLimit2,MalamentGR}; and the significance of spacetime curvature more generally \citep{Cartan1,Cartan2,Friedman}.}  We conclude that we should not have expected general relativity to have admitted of an unambiguous notion of gravitational energy in the first place.

The remainder of the paper will proceed as follows.  In the next section, we will describe the sense in which (ordinary) Newtonian gravitation admits of notions of gravitational momentum and gravitational energy.  Although we recover the expressions proposed by \citet{Synge}, we derive them via a different argument, using variational methods. In the following section, we argue that gravitational energy exhibits many of the problematic features of gravitational energy in general relativity.  We then address a potential worry---that Newtonian gravitation is saliently disanalogous, because it does not have purely gravitational degrees of freedom---by describing an analogue to the Weyl tensor in geometrized Newtonian gravitation.  We will conclude with some brief remarks about how to understand the concept of energy in light of this discussion.

\section{Gravitational Energy in Newtonian Gravitation}

We begin by analysing the situation in ordinary (i.e., non-geometrized) Newtonian gravitation. For such a theory, we assume a background classical spacetime $(M, t_a, h^{ab}, \nabla)$, where $M$ is a smooth manifold diffeomorphic to $\mathbb{R}^4$; $t_a$ and $h^{ab}$ are the standard temporal and spatial metrics on $M$;\footnote{The fields $t_a$ and $h^{ab}$ are not metrics in the strict sense of the term (i.e. neither is a symmetric nondegenerate tensor field of rank $(0,2)$); however, they do determine metrical structure on (respectively) the class of spacelike hypersurfaces and on any given spacelike hypersurface. See \cite[\S4.1]{MalamentGR} for further discussion.} and $\nabla$ is a flat derivative operator compatible with $t_a$ and $h^{ab}$ (i.e. $\nabla_a t_b = 0$ and $\nabla_a h^{bc} = 0$).\footnote{For further details on classical spacetimes, see \citet[Ch. 4]{MalamentGR}.  We assume that, in addition to $M$ being $\mathbb{R}^4$, the spacelike hypersurfaces determined by $t_a$ are all diffeomorphic to $\mathbb{R}^3$; and that $t_a$ and $h^{ab}$ are complete.} Models of the theory are of the form $(M, t_a, h^{ab}, \nabla, \rho, \varphi)$, where $\rho$ and $\varphi$ are smooth scalar fields on $M$.

In this context, the dynamical content of the theory may be specified by the following Lagrangian:\footnote{Units have been chosen so as to set $G = 1$.}
\begin{equation}\label{NGLag}
\mathcal{L} = -\rho \varphi - \frac{1}{8\pi} \nabla_a\varphi \nabla^a \varphi
\end{equation}
where $\nabla^a \varphi := h^{ab} \nabla_b \varphi$.\footnote{More generally, we will raise indices on objects defined with covariant indices by acting with the spatial metric $h^{ab}$.  Unlike in general relativity, however, we cannot in general lower indices once they have been raised, because $h^{ab}$ is not invertible.} As is well-known, if we require that for any compact set $U \subset M$, $\mathcal{L}$ is extremal with respect to variations that vanish outside $U$, then $\mathcal{L}$ satisfies the Euler-Lagrange equations. In this case, we consider variations of $\varphi$, and find that the associated Euler-Lagrange equation is Poisson's equation
\begin{equation}
\nabla_a \nabla^a \varphi = 4 \pi \rho
\end{equation}
Thus, the models of the theory correspond to the solutions to Poisson's equation.

Note that the Lagrangian in Eq. \eqref{NGLag} is invariant under temporal and spatial translations. This suggests using Noether's theorem to find associated currents, which we expect to link to notions of energy and momentum for the gravitational field. We will first articulate the general Noether argument for spacetime translations, before applying it to this specific case.\footnote{The argument below is loosely based on \cite{TongQFT}; for philosophical discussion of Noether's theorems, see \cite{BradingBrown}.} So suppose that we have some set of fields $\Psi$ over some affine space $A$ representing spacetime,\footnote{Observe that the flat derivative operator $\nabla$ on $M$, in the presence of the assumptions we have made above, makes $M$ an affine space.} and some Lagrangian $\mathcal{L} (\Psi, \nabla_a \Psi)$ with no direct dependence on spacetime: that is, we take the Lagrangian density to be a function of the field values and their first derivatives only. 
For the purposes of this analysis, it is helpful to distinguish between the Lagrangian and (what we will call) the \emph{saturated} Lagrangian $\Lambda_\Psi (x)$; the latter is a function of the spacetime manifold alone, given by
\begin{equation}
\Lambda_\Psi (x) := \mathcal{L} (\Psi(x), \nabla_a \Psi (x))
\end{equation}

Suppose that we fix a particular configuration $\Psi$ of the fields (which may or may not be a solution of the Euler-Lagrange equations for $\mathcal{L}$), and consider a constant vector field $\delta x^a$ of norm $\epsilon$.\footnote{Here we mean that if $t_a\delta x^a \neq 0$, then $|t_a\delta x^a| = \epsilon$; otherwise $h^{ab}\sigma_a\sigma_b = \epsilon^2$, where $h^{ab}\sigma_b = \delta x^a$.} For any spacetime point $x$, compare $x$ with the point $x + \delta x^a$. In the limit as $\epsilon$ approaches $0$, we have that the difference in $\Lambda_\Psi$ between these two points will be given by
\begin{equation}\label{nochain}
\delta \Lambda = \nabla_b \Lambda_\Psi \, \delta x^b
\end{equation}
However, by the chain rule, we can also express this difference in terms of $\mathcal{L}$ (since $\Lambda$ is, as it were, the result of composing $\mathcal{L}$ with $\Psi$ and its first derivative):
\begin{equation}\label{chain}
\delta\Lambda = \frac{\partial \mathcal{L}}{\partial \Psi} \delta\Psi + \frac{\partial \mathcal{L}}{\partial (\nabla_a \Psi)} \delta (\nabla_a \Psi) 
\end{equation}
where $\delta\Psi$ and $\delta (\nabla_a \Psi)$ are the differences in $\Psi$ and $\nabla_a \Psi$ between $x$ and $x + \delta x^a$. (Like Eq. \eqref{nochain}, equation Eq. \eqref{chain} is only valid in the limit as $\epsilon\rightarrow 0$.)  It follows that $\delta \Psi = \nabla_a \Psi \delta x^a$ and $\delta (\nabla_a \Psi) = \nabla_a (\delta \Psi) = \nabla_a (\nabla_b \Psi \delta x^b)$. By equating Eqs. \eqref{nochain} and \eqref{chain}, exploiting the fact that $\nabla_a \delta x^b = 0$ (since $\delta x^a$ is constant), and using the converse of the Leibniz rule, we obtain
\begin{equation}
\left[ \frac{\partial \mathcal{L}}{\partial \Psi} - \nabla_a \left( \frac{\partial\mathcal{L}}{\partial(\nabla_a \Psi)} \right) \right] \nabla_b \Psi \, \delta x^b
	= -\nabla_a \left( \frac{\partial\mathcal{L}}{\partial(\nabla_a \Psi)} \nabla_b \Psi  \right)\delta x^b
		+ (\nabla_b \Lambda_\Psi) \delta x^b
\end{equation}
Since this expression holds for any $\delta x^a$ as $\epsilon$ goes to zero, we finally have
\begin{equation}\label{noether}
\left[ \frac{\partial \mathcal{L}}{\partial \Psi} - \nabla_a \left( \frac{\partial\mathcal{L}}{\partial(\nabla_a \Psi)} \right) \right] \nabla_b \Psi = \nabla_a \tilde{T}^a{}_b
\end{equation}
where
\begin{equation}
\tilde{T}^a{}_b := -\frac{\partial\mathcal{L}}{\partial(\nabla_a \Psi)}\nabla_b \Psi + \delta^a_b \Lambda_\Psi
\end{equation}
Usually, one concludes from this that $\tilde{T}^a{}_b$ is divergence-free whenever the equations of motion are satisfied, on the grounds that those equations (i.e. the Euler-Lagrange equations) are just the assertion that the Euler operator for $\mathcal{L}$ (i.e. the term in square brackets on the left-hand side of Eq. \eqref{noether}) vanishes. However, as we are about to see, that argument can be a bit quick; so we will keep the more general expression in Eq. \eqref{noether} in play.


In order to apply the above analysis to our Lagrangian in Eq. \eqref{NGLag} for Newtonian gravitation, we need to first observe that the set of fields $\Psi$ that we used for Eq. \eqref{chain} is \emph{all} fields upon which $\mathcal{L}$ depends, not just those which we vary in applying Hamilton's principle.\footnote{See \cite[\S8]{PooleyBackground} for further discussion of non-variational fields.} Thus, in this case, $\Psi$ includes not just $\varphi$, but also $\rho$, $t_a$, and $h^{ab}$. However, it is clear that $\mathcal{L}$ only depends upon derivatives of $\varphi$, and not upon the derivatives of any of the other fields; moreover, given that $\delta x^a$ is a spatiotemporal translation, then there is no change in the spatial or temporal metrics between $x$ and $x + \delta x^a$ (i.e. $\delta t_a = 0$ and $\delta h^{ab} = 0$). So Eq. \eqref{chain} becomes
\begin{equation}
\delta \Lambda =  \frac{\partial \mathcal{L}}{\partial \varphi} \delta \varphi  + \frac{\partial \mathcal{L}}{\partial (\nabla_a \varphi)} \delta (\nabla_a \varphi) + \frac{\partial \mathcal{L}}{\partial \rho} \delta \rho
\end{equation}
Following the analysis through, we obtain
\begin{align*}
\nabla_a \tilde{T}^a{}_b &= \left[ \frac{\partial \mathcal{L}}{\partial \varphi} - \nabla_a \left( \frac{\partial\mathcal{L}}{\partial(\nabla_a \varphi)} \right) \right] \nabla_b \varphi + \frac{\partial \mathcal{L}}{\partial \rho} \nabla_b \rho\\
	&= \left[ -\rho + \frac{1}{4\pi} \nabla_a \nabla^a \varphi \right] \nabla_b \varphi - \varphi \nabla_b \rho
\end{align*}
where
\begin{align}
\tilde{T}^a{}_b &= -\frac{\partial\mathcal{L}}{\partial(\nabla_a \varphi)}\nabla_b \varphi + \delta^a_b \Lambda_\varphi\notag\\
	&= \frac{1}{4\pi} \nabla^a \varphi \nabla_b \varphi - \delta^a_b (\rho \varphi + \frac{1}{8\pi}
\nabla_n \varphi \nabla^n \varphi). \label{gravStress}\end{align}

Note that $\tilde{T}^a{}_b$ is \emph{not} divergence-free, even for solutions to Poisson's equation; instead, we find that when Poisson's equation is satisfied (i.e. when $\nabla_a \nabla^a \varphi = 4 \pi \rho$), the divergence is $\nabla_a \tilde{T}{}^a{}_b = - \varphi \nabla_b \rho$. This is because, as noted above, our Noetherian analysis must allow for variation in all the fields upon which $\mathcal{L}$ depends (not just the dynamical fields which are varied for the purposes of Hamilton's principle).  Thus, we find a current whose divergence is related to the divergence of the ``background'' field $\rho$, which in the present analysis we have not supposed to be dynamical.\footnote{There is another reason why one would not expect this current to be conserved in general: Poisson's equation only constrains the gravitational potential on a time slice; its change over time is determined by the matter dynamics (and boundary conditions).}

The expression in Eq. \eqref{gravStress} suffers from a small defect: it depends, not just on the gravitational potential, but also on $\rho$.  But we would like to isolate the energy and momentum associated with the gravitational field alone.   We can address this by subtracting away the ``interaction energy'', i.e., by observing that, since $\varphi \nabla_b \rho = \nabla_b (\rho \varphi) - \rho \nabla_b \varphi$, we may define a tensor
\begin{align*}
T^a{}_b :&= \tilde{T}{}^a{}_b + \delta^a_b \rho\varphi\\
	&= \frac{1}{4\pi} \big[ \nabla^a \varphi \nabla_b \varphi -  \tfrac{1}{2} \delta^a_b
\nabla_n \varphi \nabla^n \varphi \big]
\end{align*}
with divergence
\begin{equation}
\nabla_a T^a{}_b = \rho \nabla_b \varphi
\end{equation}
whenever Poisson's equation is satisfied.

The tensor $T^a{}_b$ in naturally interpreted as encoding the energy and momentum content of a gravitational field in Newtonian gravitation.  In particular, we take
\begin{equation}\label{NGMMT}
T^{ab}=h^{bn}T^a{}_n = \frac{1}{4\pi} \big[ \nabla^a \varphi \nabla^b \varphi -  \tfrac{1}{2} h^{ab}
\nabla_n \varphi \nabla^n \varphi \big]
\end{equation}
to be (a natural candidate for) the \emph{mass-momentum} tensor associated with the gravitational field.  And we take
\begin{equation}\label{energyDensity}
E=t_a T^a{}_b\xi^b=-\frac{1}{8\pi} \nabla_n \varphi \nabla^n \varphi,
\end{equation}
where $\xi^a$ is any unit timelike vector field, to be (a natural candidate for) the \emph{energy density} associated with the gravitational field.

In addition to the analysis given above, there are a few remarks to make about these quantities.\footnote{See also the discussion in \citet{Synge}.}  First, observe that the proposed quantity for the energy density agrees, up to a multiplicative constant, with the energy density associated with an electrostatic potential.  Given the strong formal analogies between Newtonian gravitation and electrostatics, this provides independent reasons to take Eq. \eqref{energyDensity} to define a natural candidate for the energy density of a gravitational field in Newtonian gravitation.\footnote{We observe, too, that \citet{Curiel} appears to take this quantity to be the natural candidate for the energy density of the gravitational field.}

The mass-momentum tensor is similarly well-motivated by independent considerations.\footnote{Note, however, that it also has some unusual features: most important is that it does not satisfy the so-called \emph{mass condition}, that $T^{ab}t_at_b > 0$ whenever $T^{ab}\neq \mathbf{0}$.  It follows that the gravitational field has momentum, but no mass!  Moreover, the 4-momentum, as determined by any observer, will always be spacelike, i.e., tangent to a spacelike hypersurface. (See \cite[\S4.1]{MalamentGR} for the definition of a spacelike hypersurface.)} A generic matter field whose mass density is always non-zero wherever the matter is present may always be associated with a mass-momentum tensor field with the canonical form:\footnote{When we consider ``generic matter fields'' we are working in a classical setting---and assuming, in particular, that there are no ``spin'' degrees of freedom that might require non-symmetric $T^{ab}$.}
\begin{equation}\label{MMTmatter}
T^{ab}_M = \rho \eta^a \eta^b + \sigma^{ab}
\end{equation}
where $\eta^a$ is a unit timelike field (representing the 4-velocity of the matter) and $\sigma^{ab}$ is a symmetric tensor field that is spacelike in both indices (representing the internal stress of the matter).\footnote{A discussion of mass-momentum tensors in this formulation of Newtonian gravitation is given by \citet[\S 4.1]{MalamentGR}; see also \citet{Duval+Kunzle} and \citet{DixonSR}.}  Suppose that there is some such matter interacting with a gravitational field, such that the divergence of the sum of $T_M^{ab}$ and the gravitational mass-momentum tensor vanishes, i.e. that $\nabla_a (T^{ab} + T_M^{ab}) = 0$.  It would follow that $\nabla_a T^{ab}_M = -\rho \nabla^b \varphi$. And from this, one can derive both a continuity equation $\nabla_a (\rho \eta^a) = 0$, as well as
\begin{equation}
\rho \eta^a \nabla_a \eta^b = -\rho \nabla^b \varphi - \nabla_a \sigma^{ab}
\end{equation}
which is precisely Newton's Second Law for matter subject to a gravitational force ($-\rho \nabla^b \varphi$) and internal stress forces ($-\nabla_a \sigma^{ab}$).

\section{Problems for Newtonian gravitational energy}

We have now derived candidate expressions for gravitational energy density and for a gravitational mass-momentum tensor.  To do so, we have fixed a flat derivative operator $\nabla$ on $M$ and considered scalar fields $\varphi$ that are solutions to Poisson's equation, for some mass density field $\rho$, relative to $\nabla$.  Both $\nabla$ and $\varphi$ appear in the expressions for the energy density and mass-momentum tensor that we have derived.

There is a precise sense, however, in which $\varphi$ and $\nabla$ should be thought of as ``gauge structure'' in Newtonian gravitation.\footnote{For a discussion of this idea, see \citet[\S 4.2]{MalamentGR} and \citet{WeatherallTheoreticalEquiv}; see also \citet{WeatherallUG}.  For recent discussions of how to understand Newtonian gravitation in light of this gauge structure, see \citet{Saunders}, \citet{Knox}, \citet{WeatherallMaxwell}, \citet{WallaceEmergent}, and \citet{DewarMaxwell}.}   This is because, given any pair $(\nabla,\varphi)$, where $\nabla$ is a flat derivative operator compatible with classical metrics $t_a$ and $h^{ab}$ and $\varphi$ satisfies $\nabla_n\nabla^n\varphi = 4\pi\rho$ for some smooth field $\rho$, one can always find other pairs $(\nabla',\varphi')$ such that (1) $\nabla'_n\nabla'{}^n\varphi' = 4\pi\rho$ and (2), given a curve $\gamma:I\rightarrow M$ with tangent field $\xi^a$ satisfying $\xi^a t_a = 1$,\footnote{Curves that can be parameterized in this way are called \emph{timelike}; $\xi$ is then called the unit tangent vector.}
\[
\xi^n\nabla_n\xi^a = -\nabla^a\varphi \Leftrightarrow \xi^n\nabla'_n \xi^a = -\nabla'{}^a\varphi'.\] In other words, $(\nabla',\varphi')$ will solve Poisson's equation for the same mass density field; and, since the acceleration of the center-of-mass worldline of a massive body is given by $\xi^n\nabla_n\xi^a = -\nabla^a\varphi$, where $\xi^a$ is the (normalized) tangent to that worldline, the two pairs will give rise to precisely the same trajectories for massive bodies.  Thus we have a precise sense in which $(\nabla,\varphi)$ and $(\nabla',\varphi')$ are empirically equivalent.

The sort of gauge freedom just described is pervasive in Newtonian gravitation.  Given any pair $(\nabla,\varphi)$ as above, the transformation
\begin{subequations}\label{Ngauge}
\begin{align}
\varphi 	&\mapsto \varphi' = \varphi + \psi\\
\nabla	&\mapsto \nabla' = (\nabla, C^a{}_{bc})\label{nablatrans}
\end{align}
\end{subequations}
where $C^a{}_{bc} = t_b t_c \nabla^a \psi$,\footnote{Here we are using the fact that the action of any derivative operator on $M$ may be expressed using a fixed derivative operator and a smooth tensor field $C^a{}_{bc}$. For further details, see \citet[\S1.7]{MalamentGR}.} and $\psi$ is any smooth scalar field satisfying $\nabla^a \nabla^b \psi = 0$, will generate a new pair $(\nabla',\varphi')$ satisfying (1) and (2) above. Thus, in addition to their empirical equivalence, there is a precise sense in which the models $(\nabla, \varphi)$ and $(\nabla', \varphi')$ are intertranslatable: the equations \eqref{Ngauge} specify how to interpret $\phi'$ and $\nabla'$ in terms of $\phi$, $\nabla$, and the ``gauge transformation field'' $\psi$.\footnote{This relationship is analogous to way that the gauge transformation equation for the electromagnetic potential, $A'_\mu = A_\mu - \nabla_\mu \lambda$, specifies how to interpret $A'_\mu$ in terms of $A_\mu$ and the gauge transformation field $\lambda$.}

Of course, it is controversial whether the presence of empirical-content-preserving intertranslatability suffices to warrant (or mandate) a judgment of physical equivalence. One could maintain that the models $(\nabla, \varphi)$ and $(\nabla', \varphi')$ represent empirically equivalent yet physically distinct possibilities, on the grounds that they disagree over (for example) the accelerations of bodies.\footnote{Observe that this requires taking the acceleration of a body relative to $\nabla$ in one model to be the same physical quantity as the acceleration relative to $\nabla'$ in the other model; by contrast, the translation given by equation \eqref{nablatrans} identifies the acceleration relative to $\nabla'$ with the acceleration relative to $(\nabla, C^a{}_{bc})$.} However, it does create a \emph{prima facie} case for such a judgment, for familiar reasons: if we regard these models as inequivalent, then we face the twin threats of underdetermination of model-choice by empirical facts, and of indeterminism in our theory. And certainly, insofar as empirical-content-preserving intertranslatability is considered a sufficient condition for physical equivalence in other cases (such as gauge transformations in electromagnetism, or kinematic shifts in Newtonian theories), then---we claim---it should be taken as a sufficient condition in this case also.

If we take seriously the idea that transformations of the form Eq. \eqref{Ngauge} are to be understood as gauge transformations---i.e., that they transform between physically equivalent representations of a given situation---then the quantities we derived in the previous section are problematic.  This is because neither the mass-momentum tensor defined in Eq. \eqref{NGMMT} nor the energy density defined in Eq. \eqref{energyDensity} is invariant under the Newtonian gauge transformations. In particular, the gravitational energy $E$ transforms as
\begin{equation}
E' = E - \frac{1}{4\pi} \nabla^n \psi \big( \nabla_n \varphi  + \tfrac{1}{2} \nabla_n \psi \big)\neq E.
\end{equation}
It would follow that gravitational energy in Newtonian gravitation is a gauge quantity: there are physically equivalent representations that disagree over the gravitational energy, just as there are physically equivalent representations that disagree over the electromagnetic potential, or the absolute velocity of some body. Hence, by analogy to the electromagnetic potential or absolute velocity, gravitational energy does not represent a physically meaningful quantity.

One can make essentially the same point in a different, and in some ways more insightful, way.  As noted in the introduction, there is a ``gauge-free'' reformulation of Newtonian gravitation, known as Newton-Cartan theory,\footnote{For systematic treatments of Newton-Cartan theory, see \citet{Trautman} or \citet[Ch. 4]{MalamentGR}.} on which one ``geometrizes'' the theory by passing from a pair $(\nabla,\varphi)$ to a curved derivative operator $\tilde{\nabla}$, also compatible with $t_a$ and $h^{ab}$, defined by $\tilde{\nabla} = (\nabla, \tilde{C}^a{}_{bc})$,
with $\tilde{C}^a{}_{bc} = - t_b t_c \nabla^a \varphi$.  If $(\nabla,\varphi)$ satisfies Poisson's equation for some smooth scalar field $\rho$, then this new derivative operator will satisfy the \emph{geometrized Poisson equation},
\[
\tilde{R}_{ab} = 4\pi\rho t_at_b,
\]
where $\tilde{R}_{ab}$ is the Ricci tensor associated with $\tilde{\nabla}$; likewise, $\tilde{\nabla}$ is defined so that the (timelike) curves $\gamma$ whose acceleration previously satisfied $\xi^n\nabla_n\xi^a = -\nabla^a\varphi$ will be geodesics of $\tilde{\nabla}$.  This formulation is gauge-free in the sense that any two pairs $(\nabla,\varphi)$ and $(\nabla',\varphi')$ related by Eqs. \eqref{Ngauge} will give rise to the same curved derivative operator $\tilde{\nabla}$.

In effect, $\tilde{\nabla}$ is the result of ``transforming away'' the gauge quantity $\varphi$.  But since it was precisely the gradient of $\varphi$ that entered into our candidate expressions for the mass-momentum tensor and energy density of the gravitational field, these quantities cannot be defined in the (gauge-free) models of the geometrized theory.  Concordantly, suppose that (as discussed above) we have a matter field whose mass-momentum tensor $T^{ab}_M$, as given by Eq. \eqref{MMTmatter}, satisfies $\nabla_a T^{ab}_M = -\rho \nabla^b \varphi$ with respect to the pair $(\nabla,\varphi)$.  It then follows immediately that $T_M^{ab}$ is divergence-free with respect to the curved connection, $\tilde{\nabla}_a T^{ab} = 0$. From this we obtain the equation of motion
\begin{equation}
\rho \eta^a \tilde{\nabla}_a \eta^b = - \tilde{\nabla}_a \sigma^{ab},
\end{equation}
reflecting the fact that gravitational forces have been ``absorbed'' into the curved connection. (The continuity equation is unchanged.)  It follows that there is no mass-momentum contribution from gravitation necessary to recover conservation of total mass-momentum.

So how should we understand gravitational energy from the perspective of the geometrized theory?  It is tempting to say that it is not defined: there is no quantity of gravitational energy associated with a model of geometrized Newtonian gravitation.\footnote{Of course, observing that the notion of gravitational energy already defined does not carry over to the geometrized theory does not \emph{ipso facto} rule out the possibility of defining some other notion of gravitational energy density in geometrized Newtonian gravitation.  But if such a quantity exists, it would apparently follow that it does not correspond to the natural candidate for gravitational energy density in the non-geometrized theory---and so, it would fail to satisfy a reasonable desideratum on any notion of gravitational energy in the geometrized theory.}  But this is too quick.  In fact, we \emph{can} define (many) notions of gravitational energy (and mass-momentum) in this theory, at least in some special cases.  In particular, let $(M,t_a,h^{ab},\tilde{\nabla},\rho)$ be a model of geometrized Newtonian gravitation (i.e., we assume the geometrized Poisson equation holds).  Then if the so-called \emph{Trautman curvature conditions},\footnote{See \citet[\S4.3]{MalamentGR} for a discussion of the significance of these conditions.  Note that they hold of any derivative operator $\tilde{\nabla}$ determined by a pair $(\nabla,\varphi)$ as above.} $R^a{}_b{}^c{}_d = R^c{}_d{}^a{}_b$ and $R^{ab}{}_{cd}=\mathbf{0}$ hold, it follows that we can always ``degeometrize'' the model, i.e., we can find a pair $(\nabla,\varphi)$ such that (1) $\nabla^n\nabla_n\varphi = 4\pi\rho$, and (2) for any timelike curve $\gamma$ with unit tangent vector field $\xi^a$,
\[
\xi^n\tilde{\nabla}_n\xi^a = \mathbf{0} \Leftrightarrow \xi^n\nabla_n\xi^a = -\nabla^a\varphi.
\]
Then, relative to any such degeometrization, we may define a gravitational energy density and a mass-momentum tensor.  But as we have seen, the degeometrizations of a given model of geometrized Newtonian gravitation are generally not unique: pairs related by Eqs. \eqref{Ngauge} are all degeometrizations of the same model of the geometrized theory.

Thus, we find that we may associate a gravitational energy density with a model of geometrized Newtonian gravitation only under special circumstances---e.g., when the curvature conditions are satisfied---and even then, we do so only by choosing some further structure: namely, a flat derivative operator $\nabla$, determining a class of (suitably ``flat'') inertial frames.\footnote{Here, we take a ``frame'' to be identified with a congruence of timelike curves. A derivative operator determines a class of frames: namely, the frames such that every curve in the congruence is a geodesic of the derivative operator. Conversely, given such a class of frames, a derivative operator is uniquely determined (since a derivative operator is uniquely determined by its class of geodesics).}  Likewise, we can define a mass-momentum tensor.  From this perspective, then, we can see that these quantities have an attenuated status: they are more like bookkeeping devices than anything more physical. For instance, although the mass-momentum tensor for matter is divergence-free with respect to the curved Newton-Cartan derivative operator, it will not (in general) be divergence-free with respect to the (non-unique) flat derivative operator we obtain after degeometrizing. So we introduce an artificial tensor field whose divergence will exactly cancel with the divergence of the matter mass-momentum tensor, and attribute it to ``gravity''---in exactly the same way that we introduce a ``gravitational force'' to account for the deviation of free-fall bodies from the geodesics of the flat connection.

All that said, there are some circumstances under which there is a degeometrization that is, in a certain sense, privileged or canonical.  To get at this idea, let us first consider the situation in ordinary---i.e., non-geometrized---Newtonian gravitation.  In this theory, one can consider ``island universe'' models of Newtonian gravitation, wherein matter is supported in a spatially compact region for all time.  In such models, it is natural to suppose that the gravitational potential $\varphi$ approaches zero as one approaches spatial infinity.  This assumption---or rather, stipulation---may be thought of as choosing a privileged class of inertial frames: namely, the ones for which the center of mass of the universe follows an unaccelerated trajetory.\footnote{Observe that in fact, by supposing that $\varphi$ approaches zero as one approaches spatial infinity, we are not only choosing a privileged gauge, but also specifying boundary conditions for Poisson's equation, which fixes a homogeneous solution.  (Recall that the Dirichlet problem---finding homogeneous solutions to Poisson's equation for fixed boundary conditions---has unique solutions for sufficiently well-behaved boundaries and boundary conditions \citep[Ch. 4]{FritzJohn}.)}

Of course, given such a spatially compact mass distribution $\rho$, and an associated pair $(\nabla,\varphi)$ that solves Poisson's equation and satisfies the boundary condition just described, one can pass to a curved derivative operator $\tilde{\nabla}$, just as with any model of Newtonian gravitation.  But now, there is a sense in which there is a preferred degeometrization: namely, that given by $(\nabla,\varphi)$, which will generally be the unique degeometrization for which $\varphi\rightarrow 0$ at spatial infinity.  In this case, then, it seems that there \emph{is} a well-defined, canonical notion of gravitational energy density and mass-momentum tensor: namely, that determined by the preferred degeometrization.  In fact, this sort of situation has a natural geometrical interpretation from the perspective of the geometrized theory: models of geometrized Newtonian gravitation $(M,t_a,h^{ab},\tilde{\nabla},\rho)$ with spatially compact matter distributions, admit of degeometrizations $(\nabla,\varphi)$ with the property that $\varphi\rightarrow 0$ at spatial infinity just in case they are \emph{asymptotically flat} in a certain weak and natural sense.\footnote{We will not develop this idea here; the details are given by \citet[\S4.5]{MalamentGR}.}
In other words, in the geometrized version of Newtonian gravitation, one can define a canonical gravitational energy density when a spacetime is asymptotically flat---which, recall, is precisely what is needed to define the ADM and Bondi energies in general relativity.\footnote{We do not wish to imply, however, that the situations are identical.  To the contrary, as we have described, in geometrized Newtonian gravitation, one arrives at this canonical gravitational energy via a canonical degeometrization, whereas in general relativity you arrive at it via asymptotic Killing fields that may be defined when a spacetime is asymptotically flat.  Thus one may think of the canonical gravitational energy one gets in the presence of asymptotic flatness as running through a canonical class of frames; in the relativistic case, meanwhile, the asymptotic Killing fields allows one to define a quantity that is frame-independent.  This difference may not be as significant as it looks however, since asymptotic flatness in Newton-Cartan theory could be expected to yield an analogous notion of ``asymptotic Killing field'', corresponding to vector fields whose induced one-parameter families of diffeomorphisms preserve the (curved) derivative operator at infinity.  But we will defer pursuing this idea to future work.  (Another difference, of course, is that one gets an energy density field in the Newton-Cartan case, but only an integrated global energy in general relativity.)  We are grateful to an anonymous referee for encouraging us to emphasize this point.}

Summing up, then: in general, the natural candidate for a notion of gravitational energy in ordinary Newtonian gravitation is gauge-dependent, and there is no corresponding (non-vanishing) quantity in the geometrized theory.  Still, in the presence of certain (special) boundary conditions---asymptotic flatness---one may can define a (canonical) notion of gravitational energy.  More generally (but still not in complete generality), one can define a gravitational energy density in the geometrized theory only by introducing a degeometrization.  But the resulting quantity will be frame-dependent, in the sense that it will depend on a choice of a class of frames, given by a flat derivative operator; and such a choice will generally not be physically privileged, in that many other choices would have been just as compatible with the physical structure present.
Thus energy may be defined, but it depends on extra structure, in strong analogy with general relativity.

\section{A Weyl tensor for geometrized Newtonian gravitation}

The upshot of the last section was that, although there is a sense in which one can define a notion of gravitational energy in Newtonian gravitation, the resulting quantity depends on a choice of background structure---either special boundary conditions or else a choice of frames.  Moreover, this quantity vanishes when one moves to geometrized Newtonian gravitation, suggesting that gravitational energy is not a meaningful quantity from the perspective of the geometrized theory.  All of this looks strikingly similar to the situation one encounters in general relativity.

Should this situation satisfy us, in the sense of convincing us that we should not have expected a notion of local gravitational energy in general relativity in the first place?  There is at least one reason to remain troubled.  As we noted in the Introduction, in general relativity there are rich, purely gravitational degrees of freedom that seem to ``do work'' in the sense of generating (ordinary) energy-momentum.

To give a concrete example: consider a gravitational wave propagating in the vicinity of a piezoelectric crystal, connected via a wire to a light bulb.  The crystal, when the gravitational wave passes, will be distorted.  This distortion will generate a current in the wire, which in turn will light the light bulb.  Indeed, in the LIGO experiment, the presence of a gravitational wave is detected by measuring the work done to preserve the relative position of two mirrors in an interferometer.  Thus, work in the ordinary sense is necessary to counter-act the effects of an incident gravitational wave.

One might worry that these sorts of phenomena present an important disanalogy with the Newtonian case, such that even if one does not have a satisfactory notion of gravitational energy in Newtonian gravitation, one nonetheless \emph{should} have one in general relativity.  We will now argue that this disanalogy is not as strong as it seems.\footnote{There are other arguments available to strengthen this point: for instance, there is a sense in which energy can be transfered from one system to another via a classical gravitational field: see \citet{Bondi+McCrea}, \citet{Cooperstock+Booth}, and \citet{Synge} for discussion.}

Fix a relativistic spacetime $(M,g_{ab})$.  The \emph{Weyl tensor} associated with this spacetime is defined from the metric and the curvature, as follows:
\begin{equation}
C^a{}_{bcd} = R^a{}_{bcd} - \frac{1}{2}\left(\delta^a{}_{[d}R_{c]b} + g_{b[c}R_{d]}{}^a\right) - \frac{1}{3} R \delta^a{}_{[c}g_{d]b}.
\end{equation}
The Weyl tensor has a number of striking properties.  For one, it has the same symmetries as the Riemann tensor.  It is also \emph{conformally invariant}, in the sense that given another metric $g'_{ab}$ on $M$, if $g'_{ab}=\Omega^2 g_{ab}$, then $C^a{}_{bcd}=C'^a{}_{bcd}$, where $C'{}^a{}_{bcd}$ is the Weyl tensor associated with $g'_{ab}$.  Finally, it is trace-free, in the sense that contracting the raised index with any of the lowered indices annihilates the tensor.

This last property is perhaps the most important, physically, because it permits us to decompose the Riemann tensor into the sum of terms proportional to $R_{ab}$, $R$, and the trace-free part, the last of which is not fixed by Einstein's equation.\footnote{This is not to say that Einstein's equation does not constrain the Weyl curvature: in particular, the Bianchi identities determine a relationship between the divergence of the Weyl tensor and the derivative of the Ricci tensor, which in turn is determined by Einstein's equation.}  In this way, the Weyl tensor encodes the ``purely gravitational'' degrees of freedom of general relativity, in the sense that it describes the part of the Riemann tensor that may vary even given some fixed matter distribution.\footnote{Once again (see fn. \ref{gravField}), we do not mean to say that the Weyl tensor is a candidate for a ``gravitational field'' in general relativity; rather, we mean that it captures those degrees of freedom in the theory that are not fixed by a matter distribution.  One might use the expression ``purely geometric'' instead of ``purely gravitational''.}

For instance, in the extreme case where energy-momentum vanishes everywhere, Einstein's equation implies that the Ricci tensor must vanish:
\begin{equation}
R_{ab}=\mathbf{0}.
\end{equation}
In this case, we have
\begin{equation}
C^a{}_{bcd}=R^{a}{}_{bcd},
\end{equation}
and thus the Weyl tensor completely characterizes the curvature of space-time.  The rich bestiary of vacuum solutions of Einstein's equation, then, including pure gravitational wave solutions, all correspond to different possible Weyl tensors.

This situation raises the question of whether there is anything analogous to the Weyl tensor in geometrized Newtonian gravitation.  The answer, we will argue, is ``yes,'' insofar as there are purely gravitational degrees of freedom in the theory that are represented by the part of the Riemann tensor that is not specified by the geometrized Poisson equation.\footnote{\citet{Wallace} also defines a (distinct, but closely related) Newtonian Weyl tensor and likewise associates it with homogeneous solutions to the Poisson equation; it encodes precisely the same data as the one we consider, and in this sense, it might be viewed as an equivalent proposal.  But he does not work in the geometrized formulation, nor does he emphasize or develop the analogy to the relativistic Weyl tensor.  See also \citet{Ellis+Dunsby} and \citet{Ellis} for discussions of the Weyl tensor in Newtonian theories.}  To see this, we begin by considering vacuum spacetimes---i.e., spacetimes in which $\rho=0$, and thus, by the geometrized Poisson equation, $R_{ab}=\mathbf{0}$.  In this case, any reasonable candidate for the Weyl tensor will agree with the Riemann tensor, and so we can get a sense of what a Weyl tensor would represent by studying the Riemann tensor itself.  We will then move to the more general case and present a candidate Newtonian Weyl tensor.

Fix a classical spacetime $(M,t_a,h^{ab},\nabla)$.  We will initially assume the Trautman curvature conditions, $R^{ab}{}_{cd}=\mathbf{0}$ and $R^{a}{}_b{}^c{}_d=R^c{}_d{}^a{}_b$.  (We are already assuming Poisson's equation, at least insofar as we have used it to associate vacuum solutions with vanishing Ricci curvature.)  Under these circumstances, there always exists a smooth vector field $\eta^a$ satisfying $\nabla^a  \eta^b=\mathbf{0}$.\footnote{See \citet[\S 4.2]{MalamentGR}.  The vector field $\eta^a$ satisfying these properties exists globally under the topological assumptions made in Sec. 2 above; more generally, one may always define such a vector field locally.}  Fix some such field, and let $\varphi^a$ be its acceleration field, i.e., $\varphi^a=\eta^n\nabla_n\eta^a$.  The Riemann tensor associated with $\nabla$ may then be expressed using $\varphi^a$ as:
\begin{equation}\label{vacuumCurvature}
R^a{}_{bcd}=-2t_bt_{[d}\nabla_{c]}\varphi^a.
\end{equation}
It follows from the curvature conditions that $\nabla^{[a}\varphi^{b]}=\mathbf{0}$, and thus there exists a smooth scalar field $\varphi$ such that $\varphi^a=\nabla^a\varphi$.\footnote{Again, this inference relies on our topological assumptions; more generally, it holds only locally.}  Since the Ricci tensor vanishes, it follows that $\nabla_n\nabla^n\varphi=\mathbf{0}$.  Thus, the curvature of any Ricci-flat classical spacetime satisfying the Trautman curvature conditions is fully characterized by some scalar field satisfying the homogeneous Poisson equation (relative to $\nabla$).

The converse is also true.  First, observe that whether a given scalar field $\varphi$ is a homogeneous solution to Poisson's equation does not depend on the choice of derivative operator, as long as it is compatible with $t_a$ and $h^{ab}$.\footnote{To see this, suppose we have some such derivative operator $\nabla$ and suppose $\varphi$ is such that $\nabla_n\nabla^n\varphi = 0$.  Now consider any other derivative operator $\tilde{\nabla}$ compatible with $t_a$ and $h^{ab}$.  Then $\nabla = (\tilde{\nabla},2t_{(b}\kappa_{c)}{}^a)$, where $\kappa_{ab}$ is some anti-symmetric tensor \citep[\S4.1]{MalamentGR}.  It follows that $0=\nabla_n\nabla^n\varphi = \nabla_n\tilde{\nabla}{}^n\varphi = \tilde{\nabla}_n\tilde{\nabla}{}^n\varphi + C^n{}_{nm}\tilde{\nabla}{}^m\varphi = \tilde{\nabla}_n\tilde{\nabla}{}^n\varphi$, as desired.}
Thus, we can fix some flat derivative operator $\overline{\nabla}$ and consider all homogeneous solutions $\varphi$ of Poisson's equation relative to this operator.  Then, given any such $\varphi$, there is a (unique) derivative operator $\nabla=(\overline{\nabla}, -t_b t_c\overline{\nabla}^a\varphi)$ whose associated Riemann tensor may be written $R^a{}_{bcd} =-2t_bt_{[d}\nabla_{c]}\nabla^a\varphi$.

It follows that every vacuum derivative operator $\nabla$ in geometrized Newtonian gravitation, under the assumption of the Trautman curvature conditions, has curvature that may be written as
\[
R^a{}_{bcd} = -2t_bt_{[d}\nabla_{c]}\nabla^a\varphi\]
for some homogeneous solution to Poisson's equation $\varphi$; and every homogeneous solution to Poisson's equation gives rise to a (unique) derivative operator whose curvature takes this form, for that solution.  We conclude that there are degrees of freedom of the derivative operator that are not fixed by the Ricci tensor; and that these correspond, in the presence of the Trautman curvature conditions, to different possible homogeneous solutions to Poisson's equation.  This makes good sense: these are the analog of vacuum solutions to Einstein's equation.  Moreover, two derivative operators with distinct Riemann tensors will in general have different geodesics---and so, the presence of Weyl curvature in this sense will influence the motion of bodies, much as in general relativity.  In this sense, purely gravitational degrees of freedom in geometrized Newtonian gravitation can ``do work''.

This analysis proceeded under two (strong) assumptions: one was that the Trautman curvature conditions hold of $R^a{}_{bcd}$; the other was that matter vanished.  We will now relax these conditions.

Fix, on a smooth manifold $M$,\footnote{Here we no longer make the topological assumptions of Sec. 2.} classical metrics $t_a$ and $h^{ab}$, and consider a derivative operator $\nabla$ compatible with them. We no longer suppose that $\nabla$ is a solution to Poisson's equation for a vanishing mass density $\rho$, and so we no longer suppose that the Riemann tensor and the Weyl tensor coincide; instead, as in general relativity, a candidate Newtonian Weyl tensor should be the trace-free part of the Riemann tensor, which in turn should admit a decomposition into a sum of the Weyl tensor and tensors proportional to the Ricci tensor and curvature scalar.  We will assume, however, that, whatever else is the case, $\nabla$ is spatially flat, which implies that $R^{abcd}=\mathbf{0}$, that $R^{ab}=\mathbf{0}$, and that there exists some smooth covector field $\nu_a$ such that $R_{ab}=t_{(a}\nu_{b)}$.  This assumption is reasonable for a few reasons: for one, it is really a property of the classical metrics as much as $\nabla$, in the sense that if \emph{any} derivative operator compatible with $h^{ab}$ and $t_a$ is spatially flat, then all of them are.  Moreover, in any model of Newtonian gravitation or geometrized Newtonian gravitation, the metrics always \emph{are} spatially flat, in precisely this sense, because of the geometrized Poisson equation.\footnote{Note that we do not assume that $\nabla$ satisfies the geometrized Poisson equation for any $\rho$.  Rather, we assume that whatever $\nabla$ is, it is compatible with metrics that are in turn compatible with at least one derivative operator that \emph{could} satisfy the geometrized Poisson equation, for some $\rho$ or other: it follows from this that $\nabla$ is spatially flat. See \citep[\S 4.1]{MalamentGR}.}

Given this background, we suppose that a satisfactory Newtonian Weyl tensor, $C^a{}_{bcd}$, must satisfy the following desiderata:
\begin{enumerate}
\item It should have the symmetries of the Riemann tensor, i.e., $C^a{}_{b(cd)}=\mathbf{0}$ and $C^a{}_{[bcd]}=\mathbf{0}$;
\item It should vanish under the same contractions with the classical metrics as the Riemann tensor \citep[\S 4.1]{MalamentGR}, so that $t_aC^a{}_{bcd}=\mathbf{0}$ and $C^{abcd}=\mathbf{0}$ (since we assume spatial flatness);
\item It should be trace-free;
\item It should be preserved by any diffeomorphisms that preserve the classical spacetime structure (i.e., $t_a$, $h^{ab}$, and $\nabla$); and
\item It should be at most linear in curvature tensors (i.e., it should not depend on products of the Riemann tensor with itself) and should not depend on derivatives of the Riemann tensor.\footnote{We require condition 5 because we are trying to capture the ``part'' of the Riemann tensor that is not fixed by a matter distribution, in the sense that the Riemann tensor could be written as a sum of a candidate Weyl tensor and some quantity related to the Ricci tensor, as one can do in general relativity.  Observe, too, that the relativistic Weyl tensor satisfies this condition.}
\end{enumerate}
For the next to last of these, it is sufficient that the field be constructed from $t_a$, $h^{ab}$, $R^a{}_{bcd}$, and $\delta^a{}_b$; the last condition rules out any derivatives of the Riemann tensor or higher order terms.

We note one condition that we do \emph{not} require, but which one might have imagined was important:  we do not require that the Weyl tensor be a conformal invariant.  The reason is that we are looking for a Weyl tensor for spatially flat derivative operators; in general, a conformal transformation will not preserve spatial flatness.  More importantly, this means that if $t_a$ and $h^{ab}$ are compatible with any derivative operators satisfying Poisson's equation, then in general conformally equivalent ones will not be compatible with \emph{any} derivative operator satisfying Poisson's equation for any $\rho$.  This suggests that conformal transformations just do not have any physical significance in geometrized Newtonian gravitation.  (Arguably constant conformal transformations do, but any Weyl tensor satisfying the conditions above will automatically be invariant under constant conformal transformations.)

Given this discussion, we propose the following candidate for the Newtonian Weyl tensor:\footnote{One can easily establish that all five conditions above are satisfied.  Indeed, we claim that this tensor is the unique one satisfying all five.  A sketch of the proof is that all terms proportional to $R=R_{ab}h^{ab}$, $t_{[c}t_{d]}$, or $R_{[cd]}$ vanish identically, as do terms of the form $t_bt_{[c}R_{d]n}h^{na}$.  This list exhausts the candidates to appear in an expression satisfying the desiderata above.}
\begin{equation}\label{NewtonWeyl}
C^a{}_{bcd} = R^a{}_{bcd} - \frac{2}{3} \delta^a{}_{[d}R_{c]b}
\end{equation}
This tensor captures the ``homogeneous part'' of the Riemann tensor---including sourced gravitational fields far from sources.

We conclude that (1) there are Weyl-like gravitational degrees of freedom in Newtonian gravitation, analogous to those described by the Weyl tensor in general relativity, and (2) in the presence of the Trautman curvature conditions, these are naturally identified with homogeneous solutions to Poisson's equation.  This suggests that we should take homogeneous solutions to the Poisson equation to include, among other things, the Newtonian analog of gravitational waves.\footnote{It is interesting to consider the relationship between these ``gravitational waves'' and the classical limit of gravitational plane waves discussed by \citet{EhlersLimit2}.  Briefly, the solutions we consider are homogeneous solutions to Poisson's equation at a time; we do not place any constraints on how these vary over time.  \citet{EhlersLimit2}, meanwhile, considers classical gravitational fields that are varying over time, so that at each time one has some solution to the homogeneous Poisson equation, but these smoothly vary according to a periodic function analogous to the time-dependence of a gravitational plane wave in general relativity.  In this sense, one can take the solutions that Ehlers considers to be a special case of the ones we are considering.}

Of course, there are important disanalogies here.  The Poisson equation is elliptic, and so its solutions---including its homogeneous solutions---do not propagate in the sense that solutions to hyperbolic equations do.  This means, for instance, that we cannot think of homogeneous solutions as solutions to a wave equation.  (Still, they may be ``wavy'', in the sense of exhibiting some periodicity: e.g., $\varphi=e^xe^y\sin(\sqrt{2}z)$ is a homogeneous solution to Poisson's equation that is sinusoidal in the $z$ direction.)

A closely related disanalogy is that homogeneous solutions to Poisson's equation, and thus Newtonian Weyl curvature, are wholly determined on any spacelike slice by boundary conditions at infinity on that slice.  In this sense, one can think of them as describing non-local degrees of freedom at a time; moreover, solutions at one time do not constrain solutions at subsequent times, so that one needs to implement boundary conditions at all times in order to ensure unique evolution.\footnote{\citet{Wallace} emphasizes this point in discussing Newtonian cosmology.  Note, though, that thinking of homogeneous solutions to Poisson's equation as the analogue to Weyl degrees of freedom in general relativity leads to a caveat to his analysis: although in the initial value formulation of general relativity, one does not need to specify boundary conditions at all times to get unique evolution, one does need to specify essentially independent initial data for the Weyl tensor everywhere on a Cauchy surface (i.e., specifying the Weyl tensor ``at infinity'' at a time is not sufficient).  Wallace is correct insofar as there is surely a metaphysical---and physical---difference between a situation where data at one time determines the physics everywhere, and one where boundary conditions must be specified at all times.  But there is a certain epistemological parity, nonetheless: in both cases, one must make strong assumptions about regions of space that one has no access to.}   This situation is strikingly different from that in general relativity, where boundary conditions on a particular spacelike hypersurface generally do not constrain the Weyl curvature on that hypersurface.

Still, these disanalogies make good physical sense in the setting we are considering, and do not undermine the interpretation of homogeneous solutions to the Poisson equation as the classical analogue of gravitational waves.  Indeed, we may think of Newtonian Weyl curvature as characterizing relativistic Weyl curvature---including gravitational waves---in the limit where the metric light cones flatten---and thus, the propagation velocity diverges.\footnote{For a discussion of this limit---and specifically, the sense in which geometrized Newtonian gravitation may be understood as the classical limit of general relativity---see \citet{EhlersLimit1,EhlersLimit2}, \citet{Kunzle}, \citet{MalamentLimit2}, and \citet{FletcherLimit}.}  In this limit, one would expect gravitational degrees of freedom to propagate instantaneously, in precisely the way described by an eliptic equation.

\section{Conclusion}

The considerations presented above are intended to support the following two theses: First, although gravitational energy can be defined in Newtonian gravitation theory, it turns out to be problematic---particularly in the geometrized formulation---in much the same ways as in general relativity.  In particular, although one can define a notion of gravitational energy, it depends on further structure, such as some choice of gauge---in this case, in the form of a flat derivative operator picking out a class of preferred frames---or boundary conditions that effectively provide a privileged choice of gauge.  Second, the fact that there are purely gravitational degrees of freedom in general relativity should not make the analogy just described any less satisfying---nor should it make the non-existence of (local) gravitational energy any more troubling.  In fact, geometrized Newtonian gravitation has degrees of freedom that are strongly analogous to Weyl degrees of freedom in general relativity, despite the non-existence of an unambiguous notion of gravitational energy.

The fact that gravitational energy gives rise to conceptual problems in general relativity and geometrized Newtonian gravitation suggests an important lesson for how to understand energy in geometrized theories.  In particular, there is a deep relationship between the classical notions of energy, work, force, and inertia.  Energy density is a measure of the ability to do work---i.e., to exert force over distance.\footnote{Pace \citet{Synge}, who suggests that work is an inappropriate notion in a field theory.}  But in theories in which gravitation is ``geometrized,'' in the sense that gravitation is understood as an inertial effect in curved spacetime, we should not think of gravitation as a force at all---and so, in particular, it is not the sort of thing that does work.

To the contrary, work makes sense only as a measure of the deviation from inertial motion over some distance.  This means that we should not expect there to be any gravitational energy density associated with general relativity or geometrized Newtonian gravitation---and to recover any such notion, we must introduce some background ``degeometrization,'' relative to which we can understand gravitational effects as deviations from a preferred state of motion.  Doing this involves introducing precisely the additional structure required to define gravitational energy in either theory.

This analysis raises interesting questions concerning the status of energy in other ``geometrized'' theories.  For instance, there are strong analogies between general relativity and the geometry of (classical) Yang-Mills theory, including electromagnetism.\footnote{For details, see for instance \citet{TrautmanFB} or \citet{WeatherallFBYMGR}.}  In these theories, charged matter may be represented by sections of a vector bundle associated with some principal bundle with (curved) principal connection.  The dynamics of the matter, then, depends on the curvature of the bundle, in much the same way that the dynamics of matter depends on the curvature of spacetime in general relativity.

We regularly associate a quantity of energy, or energy-momentum, with, for instance, Maxwell fields.  But there is a sense in which these fields should be viewed, much like the curvature of spacetime, as measuring something about default trajectories (for appropriately charged matter), rather than measuring a tendency to deviate from default trajectories---and thus, by the same reasoning as above, should not be understood as ``doing work''.  Does this mean that we should not associate an energy density density with Maxwell fields?  Not exactly.  But it does suggest that we should understand the energy density of Yang-Mills fields as \emph{relative} to some background structure---namely, the inertial structure determined by the spacetime metric in general relativity.  In this case, however, we have a physically privileged background relative to which we can make these determinations.

\section*{Acknowledgments}
This paper is partially based upon work supported by the National Science Foundation under Grant No. 1331126.  We are grateful to David Malament for helpful conversations about the material presented here, including suggesting the form of Eq. \eqref{NewtonWeyl}; to Erik Curiel for discussions about conformal invariance in the context of geometrized Newtonian gravitation; and to David Wallace for discussions of these and related ideas.  The manuscript was improved by comments from James Read and two anonymous referees.  Weatherall is grateful for feedback from an audience at the 18th UK/European Foundations of Physics Conference in London, UK.

\singlespacing

\end{document}